\documentclass[conference, a4paper, 10pt]{IEEEtran}
\usepackage[a4paper, left=44pt, right=44pt, bottom=72pt, top=54pt]{geometry}
\IEEEoverridecommandlockouts

\usepackage{float}
\usepackage{amsmath,amssymb}
\usepackage{graphicx}
\usepackage[short]{optidef}
\usepackage{multirow}
\usepackage{balance}

\usepackage{xcolor}

\renewcommand{\t}{^\top}
\renewcommand{\matrix}[1]{\begin{bmatrix}#1\end{bmatrix}}
\DeclareMathOperator*{\argmin}{arg\,min}

\usepackage{tikz}
\usepackage{eso-pic}

\newcommand\acceptedtext{%
  \footnotesize © 2024 IEEE. Personal use of this material is permitted. Permission from IEEE must be obtained for all other uses, in any current or future media, including reprinting/republishing this material for advertising or promotional purposes, creating new collective works, for resale or redistribution to servers or lists, or reuse of any copyrighted component of this work in other works.}

\newcommand\acceptednotice{%
  \AddToShipoutPicture* {%
    \begin{tikzpicture}[remember picture,overlay]
      \node[anchor=south,yshift=15pt] at (current page.south) {\fbox{\parbox{\dimexpr0.65\textwidth-2\fboxsep-2\fboxrule\relax}{\acceptedtext}}};
    \end{tikzpicture}%
  }%
}

\begin{document}

\bstctlcite{IEEEexample:BSTcontrol}

\title{Model Predictive Traction Control System Based on the Koopman Operator
\thanks{This work has been supported in part by the Croatian Science Foundation under the project UIP-2019-04-6487.}
}

\author{\IEEEauthorblockN{Josip Kir Hromatko}
\IEEEauthorblockA{\textit{University of Zagreb, Faculty of} \\
\textit{Electrical Engineering and Computing} \\
Zagreb, Croatia \\
josip.kir.hromatko@fer.hr}
\and
\IEEEauthorblockN{Šandor Ileš}
\IEEEauthorblockA{\textit{University of Zagreb, Faculty of} \\
\textit{Electrical Engineering and Computing} \\
Zagreb, Croatia \\
sandor.iles@fer.hr}
}

\maketitle
\acceptednotice

\begin{abstract}
Due to their importance, traction control and anti-lock braking systems have become standard equipment in modern vehicles. However, accurate models of tire dynamics are often difficult to obtain and usually include nonlinearities, making their use in control systems challenging. This paper describes a traction control system based on model predictive control and Koopman operator theory, which aims to approximate nonlinear systems with linear ones through a state space transformation. A linear model predictive controller based on the Koopman predictor is compared to a standard nonlinear model predictive controller. Experiments in a high-fidelity vehicle dynamics simulation environment show a comparable reference tracking performance of the two controllers, with a reduced execution time for the proposed Koopman operator-based algorithm, both on a standard PC and embedded hardware.
\end{abstract}


\section{Introduction}
Traction forces between the tires and the ground are the main driver of a vehicle's behaviour. Therefore, controlling these forces, which usually translates to controlling the tire slip ratio, can improve both safety and performance, as indicated by the presence of Anti-Lock Braking Systems (ABS) or traction control systems in most modern vehicles. Very often, these systems use a simple control logic (such as PID or sliding-mode controllers) to enable fast reaction times and reduce the overall costs of implementation. However, in some situations, more advanced control algorithms can provide improved performance, robustness or adaptiveness. These benefits are even more present in electric vehicles, which often allow for independent control of each wheel \cite{tc_survey}.

Among the many existing control methods, model predictive control has shown great potential, both in research and in a practical context \cite{mpc_survey}. It is based on repeatedly solving a constrained optimization problem which includes the system dynamics, a user-defined cost function and possible system constraints (e.g., on certain states or control inputs). The solution of this problem gives the optimal control inputs within a given prediction horizon. The first input is then applied to the system and the updated optimization problem is solved in the next iteration. The main benefits of model predictive control are its intuitiveness, the possibility of systematically including specific constraints and the simplicity of controlling systems with multiple inputs or outputs. On the other hand, the main drawback of this control method is the requirement for a plant model (which can sometimes be time-consuming to derive) and the computational cost of solving the optimization problem online. However, there are several methods of simplifying the modelling process and reducing the execution time of the algorithm.

Modelling of dynamical systems is most commonly done with systems of ordinary differential equations, which are often nonlinear. The most straightforward approach for controlling such systems is to use nonlinear model predictive control (NMPC), which has mild restrictions on the problem formulation, but can lead to non-convex problems. Using NMPC for traction control or as an ABS has been investigated in several papers: \cite{tc_mpc_2017} compares linear and nonlinear MPC for ABS, \cite{tc_empc} describes an explicit NMPC traction control system with a reduced computational load, \cite{abs_nmpc} presents a centralized ABS system using implicit NMPC and \cite{tc_nmpc_2022_axle} considers a per-axle traction control system using NMPC. However, despite the advances in solving nonlinear problems, real-time applicability of NMPC for traction control is still questionable.

In an attempt to enable faster execution by simplifying the MPC formulation, nonlinear models are often approximated by linear ones. With a linear system model, convex constraints and a convex cost function, the optimization problem becomes convex and can be solved much faster than a general nonlinear program. Recently, there has been an increased interest in the Koopman operator theory \cite{koopman}, which aims to approximate nonlinear systems by linear ones through a nonlinear transformation of the state space. It has been applied in system modelling, estimation and control \cite{modern_koopman_review} and vehicle dynamics control in particular \cite{koopman_vehicles_review}. Recent papers present the application of Koopman operator theory and linear MPC to active steering \cite{vit2020,vit2023}, torque vectoring \cite{svec2021,svec2023} and velocity tracking \cite{xiao2023}. However, the only work addressing control of longitudinal wheel slip with the Koopman operator theory is \cite{kmpc_andrea}, where a comparison with NMPC and a linear model-based MPC was given, showing similar performance and a reduced execution time when using the Koopman operator-based approach.

This paper presents an MPC algorithm for traction control of a single wheel, based on the Koopman operator approximated using extended dynamic mode decomposition \cite{williams2015data}. The proposed algorithm is compared with an NMPC controller in a high-fidelity simulator and on embedded hardware. The main advantage of approximating the nonlinear system with a linear model is the faster execution time, allowing for real-time implicit MPC on automotive hardware.
\section{Wheel dynamics}
A common approach in traction control system design is to consider the dynamics of a single wheel. Additionally, in this paper the vehicle's center of mass is assumed to be centered. If load transfer during acceleration is ignored, a vertical load of a quarter of the vehicle's mass $m$ can be used:
\begin{equation}
    F_z = \frac{mg}{4}
\end{equation}
The longitudinal force generated by the tire can be modelled using a simplified version of the Pacejka's formula \cite{pacejka}:
\begin{equation}\label{eq:tire_model}
    F_x = \mu_x\cdot F_z\cdot D\sin(C\arctan(B\kappa))
\end{equation}
where $\kappa$ denotes the \textit{slip ratio}, usually calculated as:
\begin{equation}
    \kappa = \frac{\omega R-v}{\omega R}
\end{equation}
in the case of vehicle acceleration. However, to improve numerical accuracy at low speeds, a modified slip definition based on \cite{micheli2023} was used:
\begin{equation}\label{eq:slip_mod}
    \kappa = \frac{(\omega R-v)\omega R}{(\omega R)^2+\epsilon},\quad \epsilon=0.1\ (\text{m/s})^2
\end{equation}

Tire model parameters $D$, $C$ and $B$ are determined by fitting the model to the tire testing data. Also, it is assumed that the road friction coefficient $\mu_x$ scales the entire tire force curve, preserving its shape and, consequently, the slip ratio corresponding to peak traction force.

If the longitudinal and rotational wheel speeds are chosen as the states, the dynamics can be expressed as:
\begin{subequations}\label{eq:vw_nl_model}
\begin{align}
    \dot{v} &= \frac{4}{m} F_x\\
    \dot{\omega} &= \frac{1}{J}(T_w-F_xR)
\end{align}
\end{subequations}
where $R$ and $J$ denote the wheel radius and moment of inertia. The torque acting on the wheel is denoted as $T_w$ and assumed to be proportional to the electric motor torque $T_m$:
\begin{equation}
    T_w = i_{gbx}\cdot T_m
\end{equation}
where $i_{gbx}$ denotes the fixed gearbox ratio. Finally, for our purposes, aerodynamics, friction and rolling resistance are assumed to have a negligible effect on wheel dynamics.

With the state vector $x=[v\ \omega]\t$ and the input $u=T_m$, the wheel dynamics \eqref{eq:vw_nl_model} can be described compactly as a continuous-time nonlinear controlled dynamical system:
\begin{equation}\label{eq:ss_cont}
    \dot{x}=f_c(x,u)
\end{equation}
with a similar formulation for its discrete-time counterpart:
\begin{equation}\label{eq:ss_discrete}
    x^+=f_d(x,u)
\end{equation}
obtained by using, e.g., a Runge-Kutta integration method.
\section{Identification of the Koopman predictor}
\subsection{Koopman operator theory for controlled systems}
The theory presented in the original paper \cite{koopman} pertained to \textit{autonomous} systems, i.e., systems of the form $x^+=f(x)$. Several papers describe extending the theory to \textit{controlled} dynamical systems \cite{proctor2018,williams2016}. The main idea is to introduce a new, extended state:
\begin{equation}
    \chi = \matrix{x\\U}
\end{equation}
where $U$ denotes an input sequence, $U=[u_0\t,u_1\t,\dots,u_\infty\t]\t$. The dynamics of the extended system can then be formulated in the autonomous form:
\begin{equation}\label{eq:koop_ext}
   \chi^+ = g(\chi)=\matrix{f(x,u_0)\\\mathcal{S}U}
\end{equation}
where $\mathcal{S}$ denotes the left shift operator, $\mathcal{S}u_i=u_{i+1}$. Finally, the associated Koopman operator $\mathcal{K}$ is defined by:
\begin{equation}
    \Phi(\chi^+) = \mathcal{K}\Phi(\chi)
\end{equation}
with $\Phi$ denoting the \textit{lifting functions}, often called \textit{observables}. Importantly, the Koopman operator is a typically infinite-dimensional \textit{linear predictor}, approximating the dynamics \eqref{eq:koop_ext} through a linear autonomous system:
\begin{equation}
    z^+ = Az
\end{equation}
where $z=\Phi(\chi)$ and $A$ is the transition matrix. By including the original states $\chi$ in the observables, the predicted states of interest can be recovered with a simple selection matrix.

\subsection{Extended dynamic mode decomposition}
EDMD was introduced in \cite{williams2015data} as a data-based method to approximate the Koopman operator using a dictionary of basis functions. For a discrete-time system of the form \eqref{eq:ss_discrete} and a dataset containing $N$ triples $(x_k,u_k,x_k^+)$, the approximation is found by solving the optimization problem:
\begin{equation}\label{eq:koop_id_ab}
    \hat{A},\hat{B}=\argmin_{A,B} \sum_{k=1}^{N} \left\|\Phi (x_k^+) - A\Phi(x_k) - B u_k \right\|_2^2
\end{equation}
where $\Phi(\cdot)$ denotes the lifting function. The states $x_k$ and $x_k^+$ are obtained from real system data or by simulating a system's nonlinear model \eqref{eq:ss_cont} for $N$ time steps. A discrete linear system approximation of the dynamics is then defined with matrices $\hat{A}$ and $\hat{B}$:
\begin{subequations}\label{eq:koop_lti}
\begin{align}
    z^+ &= \hat{A}z + \hat{B}u\\
    y &= \hat{C}z
\end{align}
\end{subequations}
with the lifted state $z=\Phi(x)$, the identified matrices $\hat{A}$, $\hat{B}$ and a selection matrix $\hat{C}$ which recovers the states of interest (i.e., the system outputs) $y$. This matrix can be set manually (e.g., if only a subset of the original states is required) or obtained by minimizing the least squares cost:
\begin{equation}\label{eq:koop_id_c}
    \hat{C} = \argmin_{C} \sum_{k=1}^{N}\left\|y_k - C \Phi(x_k) \right\|_2^2
\end{equation}

Note that both \eqref{eq:koop_id_ab} and \eqref{eq:koop_id_c} are \textit{linear} least squares problems, which can be solved analytically as described in \cite{korda2018}. Also, the resulting predictor \eqref{eq:koop_lti} minimizes the \textit{one-step} prediction, which implies that the dataset can contain multiple trajectories with as little as a single sample, i.e., the triple $(x_k,u_k,x_k^+)$. Finally, the system inputs $u_k$ enter untransformed into \eqref{eq:koop_id_ab}, which limits the prediction accuracy for systems with a nonlinear input-to-state mapping. Therefore, this method is more suitable for systems with a linear input-to-state mapping.

\subsection{Approximating the wheel dynamics}
\subsubsection{Data collection}
To identify the approximate linear system from nonlinear dynamics, an informative dataset is needed. For this purpose, we used the system model \eqref{eq:vw_nl_model} and generated 1000 trajectories of 250 samples with a sampling time of 2 ms. The initial state was determined from the random and uniformly distributed vehicle speed in the range [0 40] km/h and wheel slip ratio in the range [0 0.2]. The random motor torque was also uniformly distributed between 50 and 175 Nm.

This method was chosen instead of collecting data directly from the high-fidelity simulator in order to enable faster prototyping, but also to allow for a fair comparison between the two predictive controllers.

\subsubsection{State selection}
The choice of states for describing the traction control problem in a state-space form is not unique. Both $x=[v\ \omega]\t$, $x=[\kappa\ \omega]\t$, $x=[\kappa\ v]\t$ and $x=[s\ \omega]\t$ (where $s=\omega R-v$) can be found in the literature. We chose the latter formulation, $x=[s\ \omega]\t$, for two reasons:
\begin{itemize}
    \item rotational speed dynamics are faster than the longitudinal speed dynamics, which are approximately constant during the prediction horizon commonly used in traction control
    \item choosing $s=\omega R-v$ as the second state leads to a \textit{linear} input-output relationship of the system dynamics, $\dot{x}=f(x)+Bu$, which is desirable for the Koopman operator framework and EDMD in particular
\end{itemize}
Note that the data collected in simulation (or from a real vehicle) can be easily transformed to match the desired states. Also, in order to improve the conditioning of \eqref{eq:koop_id_ab}, the collected dataset was normalized to have zero mean and unit standard deviation.

\subsubsection{Choosing the lifting function}

In this work, we transform the original state variables into a higher-dimensional space using polynomial basis functions.  For an original state $x \in \mathbb{R}^{n_x}$, these functions are defined as:
\begin{equation} \label{eq:basis_poly}
    P_d = \left\{\prod_{j=1}^{n_x}{x_j^{v_j}} | {v_j} \in \mathbb{N} \cup \{0\}, \ \sum_{j=1}^{n_x}{v_j} \leq d \right\}
\end{equation}
where $d$ is the basis order and $v_j$ are the exponents of the state variables $x_j$.
Although different basis functions, such as Gaussian or thinplate radial basis functions, could be used for lifting the original state, the polynomial basis functions provided the highest approximation accuracy in our tests. 

Additionally, one of the benefits of formulating an MPC problem using the Koopman operator is the possibility of converting nonlinear constraints and costs into linear ones by adding the nonlinear functions in the lifted state \cite{korda2018}. In order to track the slip ratio \eqref{eq:slip_mod} using a linear MPC framework, it was added to the lifted states. The complete lifted state then becomes:
\begin{equation}\label{eq:lift}
    z=\matrix{P_d\t & \kappa}\t
\end{equation}

\subsubsection{Approximation results}
\begin{figure}
    \centering
    \includegraphics[width=\columnwidth]{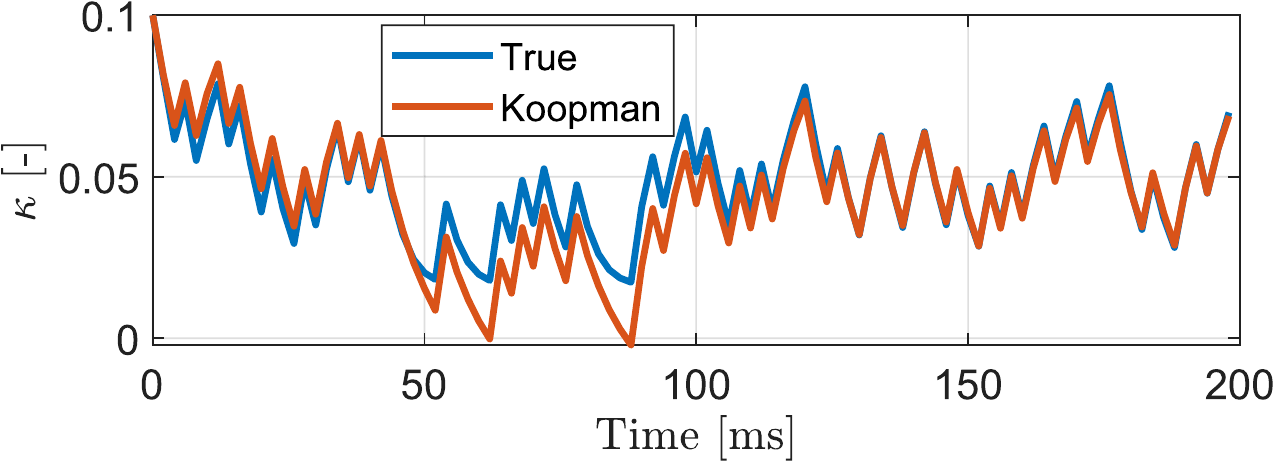}
    \caption{Prediction using the identified operator, $v_0=20$ km/h, $\kappa_0=0.1$.}
    \label{fig:sysid}
\end{figure}

Finally, the Koopman predictor \eqref{eq:koop_lti} was determined using EDMD and the polynomial basis functions \eqref{eq:basis_poly} of order 4. \figurename\ \ref{fig:sysid} shows the prediction accuracy for the initial vehicle speed of 20 km/h and PRBS control inputs in the range [50 175] Nm. Note that the time interval in the plot equals to 100 prediction steps at a sampling time of 2 milliseconds, justifying the use of this model for typical prediction horizons in traction control system design (2-10 steps).

\section{Traction control system design}
\subsection{Control goal and framework}
The main objective of a traction control system is to prevent excessive wheel slip and ensure that the optimal traction force is generated. This can usually be achieved in several ways, one of which is tracking a wheel slip setpoint. If the setpoint corresponds to the peak of the tire force curve, good traction will be obtained.

On the other hand, it is very important for driver assistance systems not to change the expected vehicle behaviour significantly. When a driver presses the acceleration pedal, they are expecting a certain vehicle acceleration or wheel torque. Therefore, the presented control algorithms are only allowed to reduce the requested motor torque (and thus prevent wheel slip). The modified torque reference is then passed on to the motor control unit, while the motor dynamics are neglected in the control algorithm.

Finally, estimating the current wheel slip can be challenging, especially if all wheels are driven. However, we are focusing on front wheel-drive electric vehicles. In that case, the longitudinal speed of the vehicle and the wheel slip can be estimated from the rear (non-driven) wheel speeds \cite{slip_est}.

\subsection{Nonlinear MPC}
With a nonlinear MPC algorithm, the nonlinear system dynamics can be used directly in the problem formulation. The cost function and the constraints can also be nonlinear, but often polytopic constraints and a quadratic cost function are used to improve convergence.

In the NMPC controller, the dynamics of the modified state $x=[s\ \omega]\t$ were modelled as:
\begin{subequations}
\begin{align}
    \dot{s} &= \left(-\frac{R^2}{J}-\frac{1}{m}\right)F_x + \frac{R}{J}i_{gbx}T_m\\
    \dot{\omega} &= \frac{1}{J}\left(i_{gbx}T_m-F_xR\right)
\end{align}
\end{subequations}
with the longitudinal force $F_x$ defined as in \eqref{eq:tire_model}. Additionally, an integral state with the following dynamics:
\begin{equation}\label{eq:e_int}
    \dot{e}_{int} = \kappa_{ref} - \kappa
\end{equation}
was introduced for improved reference tracking in the presence of disturbances, similar to \cite{tc_empc,kmpc_andrea,integral_nmpc}. With the extended state $x=[s\ \omega\ e_{int}]\t$, the system dynamics can be written compactly as:
\begin{equation}
    \dot{x} = f_{ocp}(x,u)
\end{equation}
The control input, i.e., the motor torque, was constrained to:
\begin{equation}
    0\leq T_m \leq T_{ref}
\end{equation}
where the torque reference is set by the driver or the specified maneuver. Finally, the nonlinear cost function was defined as a weighted sum of squares of the slip tracking error, the integral state and the motor torque reduction.

The NMPC optimization problem was formulated in continuous time as:
\begin{mini!}
    {u}{\int_0^T \left(w_p(\kappa_{ref}-\kappa)^2 + w_ie_{int}^2 + w_u(T_{ref}-u)^2\right)dt}{\label{eq:opt_NMPC}}{\label{eq:opt_NMPC_cost}}
    \addConstraint{
    \dot{x}=f_{ocp}(x,u)}
    \addConstraint{
    x_0 = \hat{x}(t)}
    \addConstraint{\label{eq:nmpc_uconst}
    0\leq u\leq T_{ref}}
\end{mini!}
and subsequently discretized and converted into a nonlinear program. Here, $T$ denotes the prediction horizon, $w_p$, $w_i$ and $w_u$ the scalar weighting factors (tuning parameters) and $\hat{x}(t)$ the current state estimate. The torque reference $T_{ref}$ is assumed to be constant along the horizon. The nonlinear program is then solved in every time step and the first optimal control input, $u_0^*$, is applied to the system. 

Additionally, MPC often uses a specific terminal cost or constraint to provide stability and feasibility guarantees. However, it is generally difficult to compute invariant sets and control Lyapunov functions for nonlinear systems, so the terminal ingredients were omitted in this work for simplicity.
    
\subsection{Koopman operator-based MPC}
Once the Koopman predictor \eqref{eq:koop_lti} has been identified, it can be used as the prediction model in a classic linear MPC framework. It can also be extended with additional states in the same manner as standard LTI models. In our approach, the model was extended with a discrete-time version of the integral state \eqref{eq:e_int}:
\begin{equation}
    e_{int}^+=e_{int}+T_s(\kappa_{ref}-\kappa)
\end{equation}
The slip reference, $\kappa_{ref}$, with zero dynamics:
\begin{equation}
    \kappa_{ref}^+=\kappa_{ref}
\end{equation}
was also added to the states in order to avoid an affine term in the model, i.e., to retain the form $x^+=Ax+Bu$ instead of $x^+=Ax+Bu+d$. The full state vector for KMPC thus becomes $\xi = [z\ e_{int}\ \kappa_{ref}]\t$. To allow for using the same cost function as in the NMPC approach \eqref{eq:opt_NMPC_cost}, the system outputs were selected as:
\begin{equation}
    y=\matrix{\kappa\\e_{int}\\\kappa_{ref}}
\end{equation}
Also, the negative effect of lifting (i.e., extending) the state space on the computational time can be mitigated by reformulating the extended LTI system dynamics in the so-called \textit{dense form} \cite{korda2018}:
\begin{equation}\label{eq:dense}
    \matrix{y_1\\y_2\\\vdots\\y_{N+1}} = \Gamma\xi_0 + \Theta\matrix{u_0\\u_1\\\vdots\\u_N}
\end{equation}
in which the system outputs depend only on the initial state and the sequence of inputs ($N$ denotes the prediction horizon). In this form, the size of the transition matrix does not depend on the size of the lifted state. However, this reformulation can be numerically inaccurate, especially for marginally stable systems and longer prediction horizons \cite{num_oc}.

Since the torque reference is assumed to be constant, the cost function then becomes quadratic, which (along with the linear approximated dynamics and simple input constraints) leads to a quadratic program:
\begin{mini!}
    {U}{Y\t QY + U\t RU}{\label{eq:opt_LTI}}{\label{eq:opt_cost_LTI}}
    \addConstraint{
    Y=\Gamma\hat{\xi}_0 + \Theta U}
    \addConstraint{\label{eq:kmpc_uconst}
    U \in\mathbb{U}}
\end{mini!}
where $Y$ and $U$ represent the stacked output and input vectors in \eqref{eq:dense} and $\hat{\xi}_0$ denotes the lifted state derived from the current estimate of the original state. Also, $Q$ and $R$ stand for the block diagonal matrices consisting of the weighting factors $w_p$, $w_i$ and $w_u$. Constraint \eqref{eq:kmpc_uconst} represents the stacked version of the input constraint \eqref{eq:nmpc_uconst}. Similar to the NMPC approach, which is used as the baseline, no terminal ingredients were used in the KMPC algorithm.

\begin{figure}
    \centering
    \includegraphics[width=.9\columnwidth]{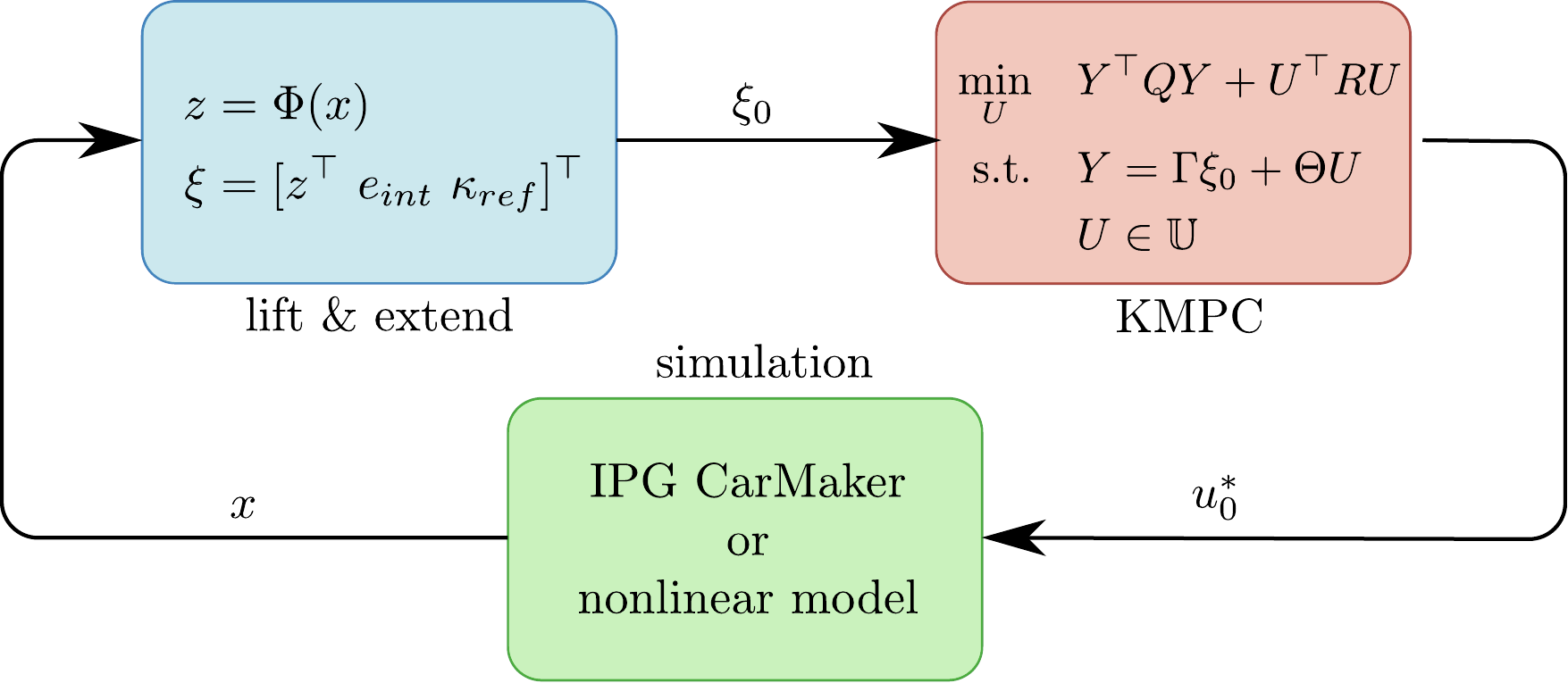}
    \caption{A schematic of the Koopman operator-based control system.}
    \label{fig:kmpc_scheme}
\end{figure}

A schematic of the Koopman operator-based MPC algorithm is shown in \figurename\ \ref{fig:kmpc_scheme}.
\section{Results and discussion}
\subsection{Test environment}
The presented predictive controllers were tested in a closed-loop simulation in IPG CarMaker, a high-fidelity vehicle dynamics simulator. The parameters of the test vehicle are given in Table \ref{tab:params}. The controllers were tested in two maneuvers, both starting from an initial speed of 2 km/h and free rolling wheels. The torque reference corresponded to simple acceleration, starting as a ramp signal and reaching the maximum torque of the motor. The slip reference for maximum traction force was estimated from the tire model and set to 0.1 in all tests.

\begin{table}
    \renewcommand{\arraystretch}{1.1}
    \centering
    \caption{Vehicle parameters}
    \begin{tabular}{c c c c}
         \hline
         Parameter & Value & Unit & Description\\
         \hline
          $m$ &   1400  & kg & mass of the vehicle\\
          $R$ &  0.318  & m & wheel radius\\
          $J$ & 1.22 & kgm\textsuperscript{2} & wheel moment of inertia\\
          $T_{max}$ & 250 & Nm & maximum motor torque\\
          $i_{gbx}$ & 3 & - & gearbox ratio\\
          $D$ & 1.2005 & - & tire model coefficient\\
          $C$ & 1.4010 & - & tire model coefficient\\
          $B$ & 17.1571 & - & tire model coefficient\\
          \hline
    \end{tabular}
    \label{tab:params}
\end{table}

Both controllers were implemented using the FORCESPRO solver \cite{forces1,forces2}, which offers a high-level interface for the NMPC formulation and an interface based on YALMIP \cite{yalmip} for the KMPC formulation. The NMPC problem was discretized using the explicit Runge-Kutta method of order 4 and solved with the nonlinear primal-dual interior point algorithm \cite{pdip_book}. Similarly, the KMPC problem was solved using a primal-dual interior point method. Controller parameters are given in Table \ref{tab:ctrl_params}, where the subscripts $(\cdot)_N$ and $(\cdot)_K$ denote the parameters related to NMPC or KMPC. Note that the difference in cost function weights is mainly caused by data scaling in the KMPC model. However, further investigating the relation between these parameters will be done in future work.

\begin{table}
    \renewcommand{\arraystretch}{1.1}
    \centering
    \caption{Controller parameters}
    \begin{tabular}{c c c }
        \hline
        Parameter & Value & Description\\
        \hline
        $T_s$ & $0.002$ s & control loop sampling time\\
        $N$ & 5 & prediction horizon length\\
        $w_{p,N}$ & $1\cdot10^3$ & NMPC slip tracking error weight\\
        $w_{i,N}$ & $1\cdot10^5$ & NMPC integral state weight\\
        $w_{u,N}$ & $1.6\cdot10^{-7}$ & NMPC torque reduction weight\\
        $w_{p,K}$ & 0.75 & KMPC slip tracking error weight\\
        $w_{i,K}$ & 500 & KMPC integral state weight\\
        $w_{u,K}$ & $1\cdot10^{-6}$ & KMPC torque reduction weight\\        
        \hline
    \end{tabular}
    \label{tab:ctrl_params}
\end{table}

\subsection{Constant road friction coefficient}
\begin{figure}
    \centering
    \includegraphics[width=.9\columnwidth]{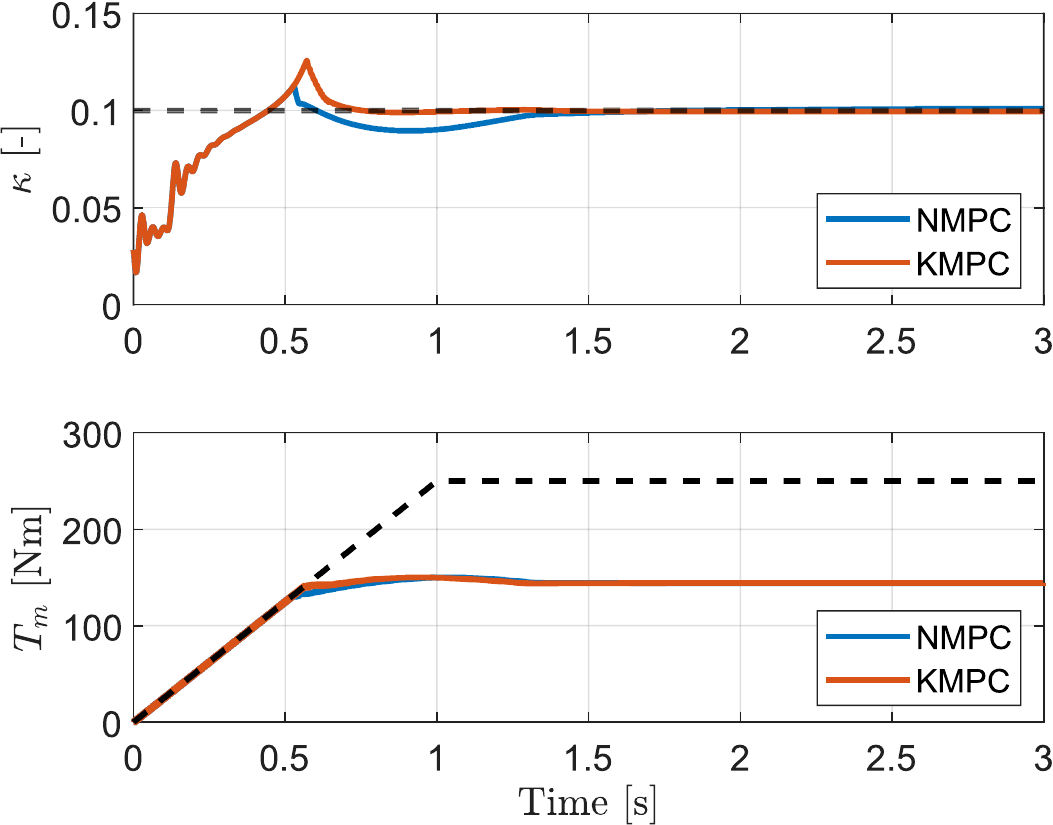}
    \caption{Simulation results with a constant road friction coefficient. Dashed lines represent reference values.}
    \label{fig:mpc_1}
\end{figure}

In the first experiment, the road friction coefficient $\mu_x$ was constant and set to 0.3. The slip tracking and torque reduction results, shown in \figurename\ \ref{fig:mpc_1}, indicate similar performance between the two controllers. After an initial ramp up of the torque and a slight overshoot of the slip reference, the torque reduction reduces the slip and improves traction.

\subsection{Varying road friction coefficient}
\begin{figure}
    \centering
    \includegraphics[width=.9\columnwidth]{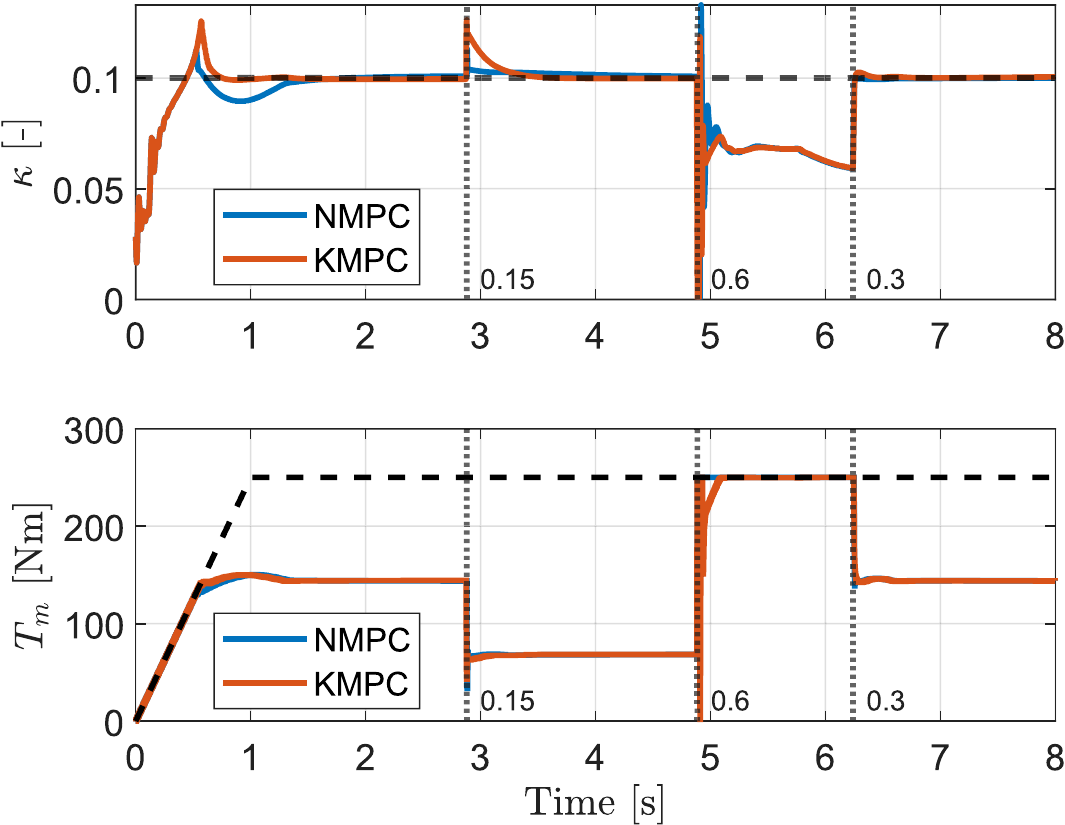}
    \caption{Simulation results with a varying road friction coefficient. Dashed lines represent reference values.}
    \label{fig:mpc_2}
\end{figure}

In the second experiment, the road friction coefficient $\mu_x$ varied along the track in the sequence 0.3-0.15-0.6-0.3. Neither controller had information about these changes and both assumed a constant road friction coefficient of 0.3. Again, the two controllers show a similar performance, with slightly larger overshoots in the case of KMPC, as shown in \figurename\ \ref{fig:mpc_2}. These overshoots occur when the road surface changes, as indicated by dashed vertical lines. Also, with a higher road surface coefficient (0.6), the slip reference cannot be reached with either controller since the motor torque is already at its maximum.

\subsection{Execution time}
Table \ref{tab:timing} shows the mean and maximum times (in milliseconds) needed to perform one controller step, i.e., to solve the optimization problem within MPC. As expected, solving a linear MPC problem (i.e., a quadratic program) is at least an order of magnitude faster that solving a nonlinear one. Importantly, this is true also for the maximum execution times, which would define the required hardware in a practical setting.

It is also important to note that the evaluation of the nonlinear lifting functions $\Phi(\cdot)$ generally also affects the time needed for calculating the optimal control input. However, it is not reported here due to the low complexity of expressions in \eqref{eq:lift}. Finally, the values presented in this section were obtained using a standard PC. Preliminary results on embedded hardware (a dSPACE MicroLabBox) align with the reported reduction in solve time, with a mean value of 1.26 ms for the NMPC and 0.023 ms for the KMPC.
\begin{table}
\centering
\caption{Execution time in milliseconds}
\begin{tabular}{c|cc|cc}
\hline
 & \multicolumn{2}{c|}{constant $\mu_x$}    & \multicolumn{2}{c}{varying $\mu_x$}     \\ \hline
controller & \multicolumn{1}{c|}{mean} & max & \multicolumn{1}{c|}{mean} & max \\ \hline
 NMPC & \multicolumn{1}{c|}{0.197} & 0.600 & \multicolumn{1}{c|}{0.210} & 0.507 \\
 KMPC & \multicolumn{1}{c|}{0.004} & 0.0432 & \multicolumn{1}{c|}{0.0037} & 0.0627 \\ \hline
\end{tabular}
\label{tab:timing}
\end{table}

\subsection{Comments}
The Koopman operator-based MPC approach has shown great potential in approximating the nonlinear MPC across many fields of application. Obtaining similar performance at lower execution times allows for a cost reduction during hardware implementation.

On the other hand, identifying the Koopman predictor is not straightforward, as noted in the literature. There are several data-driven methods, each with their own benefits and caveats. Additionally, intuitive modifications to the MPC algorithm such as extending the prediction horizon or increasing the size of the lifting function can sometimes result in the opposite of what would be expected. It is still unclear how much of the recent success in combining Koopman operator theory and MPC is due to the approximation capabilities of linear operators, as compared to the robustness of MPC to model-plant mismatch \cite{kaiser2020}.

Finally, since most of the methods for identifying the Koopman predictor rely only on data, it should be possible to automate and simplify the identification process. With the available computing power and the development of software related to the Koopman operator theory, the effort needed for modelling and controlling nonlinear systems could be significantly reduced.
\section{Conclusion}
This paper presents an approach to traction control system design for an electric vehicle based on the Koopman operator theory and model predictive control. Using the data collected by standard automotive equipment and a relatively simple predictor identification method, it was possible to obtain good approximation results with a nonlinear lifting function and a linear operator. The obtained linear system model was then used in a model predictive control framework and compared to a nonlinear model predictive controller. Simulations in a high-fidelity simulation environment showed comparable performance of the two controllers, with a reduction in execution time for the proposed algorithm.

Since the identified predictor is only an approximation of the infinite-dimensional operator, an interesting topic for future work could be to include the estimated approximation error in the linear model as a disturbance. Another aspect to consider is updating the model online, e.g., in case of a change in the tire characteristics. Finally, deploying the presented algorithm on embedded hardware could provide valuable insights about its practical applicability.

\section*{Acknowledgement}
The authors would like to thank Marko Švec for helpful discussions on various aspects of Koopman operator theory and its application to vehicle dynamics. Also, Josip Kir Hromatko would like to thank Andrea Sassella for specific suggestions related to braking systems and traction control applications.

\balance
\bibliographystyle{IEEEtran}
\bibliography{bibliography}

\end{document}